\documentclass[journal]{IEEEtran}

\usepackage[cmex10]{amsmath}
\usepackage{amssymb,mathtools,bm}
\usepackage{siunitx}
\usepackage{graphicx}
\usepackage{booktabs}
\usepackage{enumitem}
\usepackage{algorithm}
\usepackage{algorithmic}
\usepackage{cite}
\usepackage[compatibility=false]{caption}
\usepackage{subcaption}

\newcommand{\e}{\mathbf{e}}
\newcommand{\f}{\mathbf{f}}
\newcommand{\g}{\mathbf{g}}

\newcommand{\p}{\mathbf{p}}

\newcommand{\s}{\mathbf{s}}

\newcommand{\w}{\mathbf{w}}
\newcommand{\x}{\mathbf{x}}

\newcommand{\C}{\mathbf{C}}

\renewcommand{\H}{\mathbf{H}}
\newcommand{\I}{\mathbf{I}}
\newcommand{\J}{\mathbf{J}}

\newcommand{\N}{\mathbf{N}}

\renewcommand{\P}{\mathbf{P}}

\newcommand{\T}{\mathbf{T}}

\newcommand{\V}{\mathbf{V}}

\newcommand{\Y}{\mathbf{Y}}

\newcommand{\Compl}{\mbox{$\mathbb{C}$}}

\newcommand{\diag}{\mathrm{diag}}

\begin{document}
\title{RIS-Enabled Spoofing Against Adversary Sensing:
CRB-Maximizing Design and Decoying Analysis}

\author{\IEEEauthorblockN{
Ioannis Gavras$^1$, Giuseppe Thadeu Freitas de Abreu$^2$, and George C. Alexandropoulos$^{1}$ 
} 
\\
\IEEEauthorblockA{$^1$Department of Informatics and Telecommunications, National and Kapodistrian University of Athens, Greece}
\IEEEauthorblockA{$^2$School
of Computer Science and Engineering, Constructor University Bremen, Germany}
\\
\IEEEauthorblockA{emails: \{giannisgav, alexandg\}@di.uoa.gr, gabreu@constructor.university}
\vspace{-1.1cm} 
\thanks{This work has been supported by the SNS JU project TERRAMETA under the EU's Horizon Europe research and innovation programme under Grant Agreement number 101097101, including also top-up funding by UKRI under the UK government's Horizon Europe funding guarantee.}}

\maketitle

\begin{abstract}
This paper studies the capability of a Reconfigurable Intelligent Surface (RIS), when transparently covering a User Equipment (UE), to deceive an adversary monostatic radar system. A compact RIS kernel model that explicitly links the radar's angular response to the RIS phase profile is introduced, which enables an analytical investigation of the Angle of Arrival (AoA) estimation accuracy with respect to the kernel's power. This model is also leveraged to formulate an RIS-based spoofing design with the dual objective to enforce strict nulls around the UE's true reflection AoA as well as maximize the channel gain toward a decoy direction. In addition, the RIS's deception capability is quantified using point-wise and angle-range robust criteria, and a configuration-independent placement score guiding decoy selection is proposed. The presented numerical results confirm deep nulls at the true reflection AoA together with a pronounced decoy peak, a fact that eventually renders maximum-likelihood sensing at the adversary radar unreliable.
\end{abstract}


%replace everywhere \boldsymbol{w} with \mathbf{w}
\section{Introduction}
Reconfigurable Intelligent Surfaces (RISs) are engineered metasurfaces whose constituent subwavelength-space elements impose programmable phase shifts on incident waves~\cite{huang2019reconfigurable,9367575}, enabling, when placed accordingly~\cite{alexandropoulos2023ris}, a wide variety of performance improvements, range channel shaping~\cite{alexandropoulos2021reconfigurable} and massive user access~\cite{RIS_massive_access} to seamless non Line-of-Sight (LoS) localization and sensing~\cite{RIS_smart_cities}. Within the European Telecommunications Standards Institute (ETSI), a dedicated industry specification group~\cite{liu2025sustainable} is currently issuing detailed reports on use cases, single- and multi-functional architectures, orchestration protocols, and field trials, paving the way for the technology's consideration in the future 3GPP releases \cite{3gppRel20}. % need more recent release, ie, R20

In the context of the emerging paradigm of Integrated Sensing And Communications (ISAC), and, particularly, in the respective concerns for sensing security and privacy~\cite{Secure_ISAC}, RISs have been lately deployed for sensing protection against adversary radars. In~\cite{zheng2024intelligent}, RIS-aided stealth was studied with the goal to minimize the echo power at hostile Receivers (RXs), presenting, under idealized LoS assumptions, closed-form designs for a single monostatic radar as well as low-complexity solvers based on the mean square error criterion for multi-radar scenarios. The authors in~\cite{wang2024intelligent} utilized a RIS to suppress returns along the true direction from a malicious monostatic radar while redirecting energy toward clutter for the radar to infer a false Angle-of-Arrival (AoA). A hybrid RIS design was deployed in~\cite{magbool2025hiding} to shield a region from an adversary while preserving communications and legitimate sensing. In particular, an alternating optimization strategy to jointly configure the transmit and RIS-induced beamforming was presented. Very recently, in~\cite{RIS_blinding}, an RIS was optimized to occultate user positions from wiretappers, while maintaining both sensing and desired communication performance between the user and a legitimate Base Station (BS). However, the state of the art either relies on clutter to mislead an adversary from sensing or suppresses echo power, while there is a lack of a Cram\'{e}r-Rao Bound (CRB) analysis quantifying the fundamental deception limits of RIS-based spoofing.

Motivated by the RIS capability for dynamically programmable reflections and the aforedescribed research gap, in this paper, we study the potential of an RIS coating a User Equipment (UE) to camouflage the latter against an adversary monostatic radar system. In particular, an RIS-enabled spoofing design that suppresses the true echo from the UE while simultaneously synthesizing a reflection at a prescribed decoy direction is presented. A compact RIS kernel that maps the metasurface phase profile to the radar's angular response is introduced, and it is shown, via Fisher Information (FI) and CRB analyses, that reflection AoA estimation accuracy at the adversary radar scales with kernel power. The deception capability of the proposed spoofing design is characterized via point-wise and AoA-range-robust criteria, and a configuration-agnostic score to guide decoy AoA selection is introduced. Numerical results confirm that the proposed RIS-enabled spoofing implements deep nulls at the true AoA together with a pronounced decoy peak, with the latter rendering Maximum Likelihood (ML) AoA estimation deceiving.  
%and delineating the practical and theoretical limits of phase-only control RIS camouflage.

\section{System and Channel Models}
We consider a narrowband wireless system setup comprising an adversary monostatic \(N\)-antenna radar and a UE within the radar's vicinity covered entirely by an \(M\)-element RIS, whose presence is unknown to the former. The radar follows a typical sensing procedure according to which, it performs UE localization via processing the received reflections of its transmitted pilot symbols. On the other hand, the RIS is optimized to obfuscate its (equivalently, UE's) location, by synthesizing a deceptive scattering response at a prescribed decoy position, thereby, inducing false target parameter estimation, and thus, masking the UE. In this initial version of the RIS spoofing framework in this paper, we consider a $2$-Dimensional (2D) geometry with half-wavelength-spaced Uniform Linear Arrays (ULAs) for both the adversary BS and the RIS. The extension to a planar RIS is left for future work.  %RIS as a ULA is very risky

\subsection{Received Reflection Signal Model}
Let the monostatic radar BS transmit the pilot sequence \(\s\triangleq[s_0,\ldots,s_{T-1}]^{\rm T}\!\in\!\Compl^{T\times 1}\) with \(\frac{1}{T}\|\s\|^{2}=1\), after being digitally precoded with the unit-norm vector \(\f_{\rm TX}\!\in\!\Compl^{N\times1}\). The resulting baseband observation at the BS's antenna elements across \(T\) pilot symbols, \(\Y\in\Compl^{N\times T}\), is mathematically expressed as:
\begin{equation}\label{eq: signal}
    \Y\triangleq\sqrt{P}\,\H_{\rm e2e}\,\f_{\rm TX}\,\s + \N,
\end{equation}
where $\mathbf{N} \triangleq [\mathbf{n}(1),\ldots,\mathbf{n}(T)] \in \Compl^{N\times T}$, with each $\mathbf{n}(t)\sim\mathcal{CN}(\mathbf{0}_{N},\sigma^2\mathbf{I}_{N})$, represents the Additive White Gaussian Noise (AWGN) over all pilot transmissions, $P$ is the total transmit power, and \(\H_{\rm e2e}\in\Compl^{N\times N}\) is the round trip cascaded channel between the monostatic BS and the UE, which is assumed fully masked by the RIS. This assumption implies that any reflection will be impacted by the RIS phase configuration.
%\footnote{We assume that \(\H_{\text{casc}}\) contains only the downlink component reflected by the RIS; the legitimate UE, although co-located with the RIS, is fully masked and produces no direct echo at the monostatic radar, so the RIS acts as the sole scatterer in the scene. Note, that the monostatic radar is unaware of the RIS's presence and attributes any echo to the UE alone.}.

Let $\boldsymbol{\omega}\triangleq[e^{j\omega_0},\dots,e^{j\omega_{M-1}}]^T\in\Compl^{M\times 1}$ contain the tunable responses of the RIS elements and $\boldsymbol{\Omega}\triangleq\diag(\boldsymbol{\omega})\in\Compl^{M\times M}$ be the corresponding RIS configuration matrix. Under the far-field assumption, the RIS maps the incident plane wave to a re-radiated plane wave through the following scalar response~\cite{huang2019reconfigurable}:
\begin{align}\label{eq: beta}
    \beta(\psi_{\text{out}},\psi_{\text{in}};\boldsymbol{\omega})
    &\triangleq  \boldsymbol{\alpha}_M^H(\psi_{\text{out}})\,\boldsymbol{\Omega}\, \boldsymbol{\alpha}_M(\psi_{\text{in}})\nonumber\\
    &= \sum_{m=0}^{M-1} \e^{\jmath\omega_m}\,\left[\boldsymbol{\alpha}_M(\psi_{\text{out}})\right]_m^*\,\left[\boldsymbol{\alpha}_M(\psi_{\text{in}})\right]_m,
\end{align}
where $\psi_{\text{in}}$ and $\psi_{\text{out}}$ are the AoA and Angle of Departure (AoD) at and from the RIS, respectively. For an $L$-element ULA with half-wavelength antenna spacing~\cite{6184250}, the steering vector of the latter \textit{kernel} at an azimuth angle $\theta$ is given by:
\begin{equation}\label{eq: steer}
    \boldsymbol{\alpha}_L(\theta)\triangleq\left[1, e^{\jmath\pi\sin\theta}, \ldots, e^{\jmath\pi(L-1)\sin\theta}\right]^T\in\Compl^{L\times 1}.
\end{equation}
It is often convenient to reformulate $\eqref{eq: beta}$ by the RIS per-element kernel vector, which can be computed as follows: 
\begin{equation}
    \boldsymbol{\upsilon}(\psi_{\text{out}},\psi_{\text{in}})\triangleq
    \boldsymbol{\alpha}_M^*(\psi_{\text{out}})\,\odot\,\boldsymbol{\alpha}_M(\psi_{\text{in}}) \in\Compl^{M\times 1},
    \label{eq:vvec}
\end{equation}
where $\odot$ is the Hadamard (i.e., element-wise) vector product. Clearly, it holds that $\beta(\psi_{\text{out}},\psi_{\text{in}};\boldsymbol{\omega})=\boldsymbol{\upsilon}^{\rm H}(\psi_{\text{out}},\psi_{\text{in}})\,\boldsymbol{\omega}$.

\subsection{Channel Model}
Let $\psi_{\text{true}}$ and $\theta_{\text{true}}$ represent respectively the AoA of the BS signal at the RIS (i.e., the UE) and the AoA of the reflected signal at the BS. Assuming that the BS is placed at the origin, while the RIS %strictly speaking this is not a panel, it's just a line.
is at the point \(\p_{\rm RIS}\triangleq[p_x,p_y]\), the round trip cascaded channel in~\eqref{eq: signal} is modeled as follows:
\begin{equation}
    \H_{\text{e2e}}= a_{\text{RIS}}\boldsymbol{\alpha}_N(\theta_{\text{true}})\,\beta\!\big(\psi_{\text{out}},\psi_{\text{in}};\boldsymbol{\omega}\big)\,\boldsymbol{\alpha}_N^H(\psi_{\text{true}}),
    \label{eq:casc}
\end{equation}
where $a_{\text{RIS}}=\Big(\tfrac{\lambda}{4\pi\|\mathbf{p}_{\mathrm{RIS}}\|}\Big)^2$ denotes the end-to-end channel attenuation with $\lambda$ being the wavelength, and $\boldsymbol{\alpha}_N(\cdot)$ is defined similar to~\eqref{eq: steer} indicating the BS's steering vector. It must hold that $\theta_{\text{true}}=\psi_{\text{true}}=\psi_{\text{in}}={\rm atan2}(p_y/p_x)$ and $\psi_{\text{out}}=\theta_{\text{true}}+\pi$.
%anoter investigation is to model the RIS/UE as extended targets with the UE's discretization having only a portion of its extended targets being the RIS element. We can also study the portion of the UE needed to be covered by an RIS for spoofing 

\section{Proposed RIS-Enabled Spoofing Design}\label{Sec: 3}
In this section, we first derive the CRB for the AoA estimation of the reflected signal from the RIS-coated UE at the adversary monostatic radar BS. Inspired by this bound's scaling behavior, a novel optimization framework that jointly nulls the backscattered signal toward the BS to conceal the UE and synthesizes a decoy echo at a prescribed AoA to mislead the respective estimation is presented.

\subsection{CRB Analysis and Discussion}
It holds from~\eqref{eq: signal}'s inspection that, for any hypothesized AoA $\theta$, the received signal at the outputs of the BS's antenna elements modeled via \(\Y\) has the mean value \(E[\Y]= \sqrt{P}\g(\theta;\boldsymbol{\omega})\s\) and  the variance \(E[(\Y-E[\Y])(\Y-E[\Y])^{\rm H}]=\sigma^2\I_{N} \), where $\g(\theta;\boldsymbol{\omega})\triangleq a_{\text{RIS}}\bar{\beta}(\theta;\boldsymbol{\omega})\boldsymbol{\alpha}_N(\theta)v(\theta)$ is the composite scalar gain comprising of this RIS kernel $\bar{\beta}(\theta;\boldsymbol{\omega})\triangleq \beta(\theta+\pi,\theta;\boldsymbol\omega)$ and %$\u(\theta)\triangleq\boldsymbol{\alpha}_N(\theta)$, and
$v(\theta)\triangleq\boldsymbol\alpha_N^{\rm H}(\theta)\mathbf{f}_{\mathrm{TX}}$,
are the array responses. According to \cite{kay1993statistical}, the scalar FI with respect to the unknown AoA parameter $\theta$ can be computed as follows:
\begin{align}\label{eq: FI}
J(\theta;\boldsymbol{\omega})=\frac{2}{\sigma^2}\left\|\sqrt{P}\frac{\partial \g(\theta;\boldsymbol{\omega})}{\partial\theta}\mathbf{s}\right\|^2
= \frac{2PT}{\sigma^2}\left\|\frac{\partial \g(\theta;\boldsymbol{\omega})}{\partial\theta}\right\|^2
\end{align}
with
\begin{align}
\frac{\partial \g(\theta;\boldsymbol{\omega})}{\partial\theta}=&a_{\text{RIS}}\Bigg[\frac{\partial\bar{\beta}(\theta;\boldsymbol{\omega})}{\partial\theta}\,\boldsymbol{\alpha}_N(\theta)v(\theta) \\&+\bar{\beta}(\theta;\boldsymbol{\omega})\left(\frac{\partial \boldsymbol{\alpha}_N(\theta)}{\partial\theta}v(\theta)+\boldsymbol{\alpha}_N(\theta)\frac{\partial v(\theta)}{\partial\theta}\right)\Bigg].\nonumber
\end{align} 
Note that, under standard array approximations, the RIS kernel changes only weakly with angle; much more slowly than the BS array response. Thus, in a small neighborhood around the actual AoA $\theta$, the RIS kernel is typically flat, i.e., $\frac{\partial\beta(\theta;\boldsymbol{\omega})}{\partial\theta}\approx 0$ \cite{van2002optimum},
%It is noted that, in a small neighborhood around the actual AoA $\theta$, the RIS kernel is typically flat, i.e., $\frac{\partial\beta(\theta;\boldsymbol{\omega})}{\partial\theta}\approx 0$ \cite{van2002optimum}. %not clear enough
indicating that the scalar FI expression in~\eqref{eq: FI} can be approximated as follows:
\begin{align}\nonumber
J(\theta;\boldsymbol{\omega})\!\approx\!
\frac{2PT}{\sigma^2}a_{\text{RIS}}^2|\bar{\beta}(\theta;\boldsymbol{\omega})|^2\!
\left\|\frac{\partial \boldsymbol{\alpha}_N(\theta)}{\partial\theta}v(\theta)\!+\!\boldsymbol{\alpha}_N(\theta)\frac{\partial v(\theta)}{\partial\theta}\right\|^2\!\!.
\end{align}

To minimize the CRB, and hence the Position Error Bound (PEB), the BS precoding vector needs to be set to maximum-ratio transmission, i.e., $\f_{\rm TX}=\boldsymbol\alpha_N(\theta)/\sqrt{N}$, yielding the following simplification for the CRB ($\text{CRB}(\theta;\boldsymbol{\omega})\triangleq J^{-1}(\theta;\boldsymbol{\omega})$):
\begin{align}\label{eq: closed_form}
\text{CRB}(\theta;\boldsymbol{\omega})
\approx
\frac{\sigma^2}{2PTa^2_{\text{RIS}}|\bar{\beta}(\theta;\boldsymbol{\omega})|^2
\pi^2 \cos^2(\theta)N^2(N-1)^2},
\end{align}
which further results in the corresponding\footnote{It is noted that to map the angular CRB in \eqref{eq: closed_form} to the position domain \(\boldsymbol{\xi}\triangleq[\xi_x,\xi_y]^{\top}\), we apply the FI change of variables $\mathbf{J}(\boldsymbol{\xi};\boldsymbol{\omega})=\T^{\rm T} J(\theta;\boldsymbol{\omega})\T$ with $\T \triangleq \frac{\partial \theta}{\partial \boldsymbol{\xi}}$ and define the position CRB as well as the respective the PEB as
\(\text{CRB}(\boldsymbol{\xi};\boldsymbol{\omega}) \triangleq \text{Tr}\{\J(\boldsymbol{\xi};\boldsymbol{\omega})^{-1}\}\) \cite{shen2010fundamental}.} $\text{PEB}(\theta;\boldsymbol{\omega})\triangleq\sqrt{\text{CRB}(\theta;\boldsymbol{\omega})}$. This indicates that the FI expression and thus the CRB and PEB are proportional to the RIS kernel:
\begin{align}\label{eq: prop}
J(\theta;\boldsymbol{\omega}) \propto \left|\bar{\beta}(\theta;\boldsymbol{\omega})\right|^2,\qquad 
\text{CRB}(\theta;\boldsymbol{\omega}) \propto \frac{1}{\left|\bar{\beta}(\theta;\boldsymbol{\omega})\right|^2}.
\end{align}
Consequently, according to \eqref{eq: closed_form} and \eqref{eq: prop}, if the RIS imposes $\bar{\beta}(\theta_{\text{true}};\boldsymbol{\omega})\approx 0$, then $J(\theta_{\text{true}};\boldsymbol{\omega})\approx0$, implying that the CRB for $\theta_{\text{true}}$ (i.e., the error variance of the radar BS estimator) diverges and the PEB becomes unbounded at the true AoA. Conversely, concentrating the RIS kernel power at a decoy AoA $\theta_{\text{fake}}$ yields small CRB around that wrong direction.

\subsection{Problem Formulation and Solution}\label{Sec: Problem}
This paper's spoofing objective is to design the RIS so as to enforce an erroneous UE position estimation at the radar BS. In particular, the goal is to null the radar's response at the true AoA $\theta_{\text{true}}$ and direct the incident radar wave toward a decoy angle $\theta_{\text{fake}}$. Using \eqref{eq:vvec}, we define the decoy kernel vector $\w\triangleq\boldsymbol{\upsilon}(\theta_{\text{fake}},\theta_{\text{true}})$ as well as the vector with the nulling angles $\boldsymbol{\Theta}_{\text{null}}\triangleq\{\theta_1,\ldots,\theta_K\}$, where $\theta_k\in[\theta_{\text{true}}-\Delta,\theta_{\text{true}}+\Delta]$ $\forall k=1,\ldots,K$ with $\Delta$ being a width that specifies the RIS-centered angle nulling window over which we will enforce $\bar{\beta}(\theta_k;\boldsymbol{\omega})\approx0$ $\forall k$. We next collect the RIS kernel vectors associated with the angle nulling window\footnote{We impose a null over an angular range around $\theta_{\text{true}}$ rather than at a single point because point nulls can be inaccurate. Even with accurate AoA knowledge at the RIS, the existence of noise can shift the radar's response by a small offset. A banded, robust null ensures low radar response, and thus UE AoA concealment, throughout a neighborhood of $\theta_{\text{true}}$ \cite{cox1987robust}.} as follows:
\begin{equation}\nonumber
    \V \triangleq \big[\,\boldsymbol{\upsilon}(\theta_1,\theta_{\text{true}}),\ldots,\boldsymbol{\upsilon}(\theta_K, \theta_{\text{true}})\,\big]\in\C^{M\times K}.
\end{equation}

Building on \eqref{eq: FI} and the closed-form CRB expression in \eqref{eq: closed_form}, we define the FI-inspired RIS-enabled spoofing objective as:
\begin{align}\nonumber
    \mathcal{P}_1:\,\max_{\boldsymbol{\omega}}\, 
    \frac{J(\theta_{\text{fake}};\boldsymbol{\omega})}
         {\varepsilon + \sum_{k=1}^K J(\theta_{k};\boldsymbol{\omega})}\,\,\text{s.t.}\, |[\boldsymbol{\omega}]_m|=1\,\forall m,
\end{align}
where $\varepsilon>0$ is a small positive constant that regularizes the fraction and prevents divergence. By invoking the proportionality in \eqref{eq: prop}, $\mathcal{P}_1$ can be approximated as:
\begin{align}\nonumber
    \mathcal{P}_2:\,\max_{\boldsymbol{\omega}}\, 
    \frac{|\w^{\rm H}\boldsymbol{\omega}|^2}
         {\varepsilon + \sum_{k=1}^K|\boldsymbol{\upsilon}(\theta_k,\theta_{\text{true}})^H \boldsymbol{\omega}|^2 }\,\,\text{s.t.}\, |[\boldsymbol{\omega}]_m|=1\,\forall m.
\end{align}
Maximizing this ratio drives the leakage at $\boldsymbol{\Theta}_{\text{null}}$ to zero when feasible; otherwise, it trades off decoy angle gain against nulling efficiency over $\Theta_{\text{null}}$. %residual leakage?? 
Following \cite{russell2020principles,bauschke1996projection,censor2012effectiveness,dinkelbach1967nonlinear}, we recast \(\mathcal{P}_1\) under the assumptions: \textit{i}) non-aliased, distinct concealment samples (i.e., $\operatorname{rank}(\V)=K$); \textit{ii}) $M\!\ge\!2K$; and \textit{iii}) the decoy AoA is not included in $\Theta_{\text{null}}$ (i.e., $\w\notin\text{span}(\V)$). Under these conditions a feasible solution exists and, as $\varepsilon\!\to\!0^+$ and $M\to\infty$, $\mathcal{P}_1$ converges to the same optimum value with\footnote{If the hard-null constraint \(\V^{\rm H}\boldsymbol{\omega}=\boldsymbol{0}_{K\times 1}\) is feasible, the denominator of \(\mathcal{P}_1\) collapses to \(\varepsilon\) with $\mathcal{S}$, hence, maximizing \(\mathcal{P}_1\) is equivalent to maximizing \(|\w^{H}\boldsymbol{\omega}|^{2}\), i.e., \(\mathcal{P}_3\) up to a global phase configuration. On the other hand, when exact nulling is infeasible, \(\mathcal{P}_1\) yields a smooth decoy-leakage trade-off and closely approximates the hard-null solution~\cite{boyd2004convex}.}:
\begin{align}
    \mathcal{P}_3:\,\max_{\boldsymbol{\omega}}\quad  |\w^{\rm H}\boldsymbol{\omega}|\,\,\text{s.t.}\, \V^{\rm H}\boldsymbol{\omega}=\boldsymbol{0}_{K\times 1},\,|[\boldsymbol{\omega}]_m|=1\,\forall m. \nonumber
\end{align}
The latter optimization problem is a quadratic program over a nonconvex feasible set imposed by linear equality constraints and unit-modulus entries, making direct solution impractical. We therefore opt for a lightweight alternating projections procedure. To this end, we define the following sets:
\begin{align}
&\nonumber\mathcal{S} \triangleq \{\boldsymbol{\omega}\in\mathbb{C}^{M\times 1}:\ \V^{\rm H}\boldsymbol{\omega}=\boldsymbol{0}\},
\\
&\nonumber\mathcal{T} \triangleq \{\boldsymbol{\omega}\in\mathbb{C}^{M\times 1}:\ |[\boldsymbol{\omega}]_m|=1\, \forall m\},    
\end{align}
implying that $\mathcal{P}_3$'s feasible solution set is their intersection: \(\mathcal{F}\triangleq\mathcal{S}\cap\mathcal{T}\).
Assuming \(\V\) has full column rank, the orthogonal projector onto \(\mathcal{S}\) is defined as:
\[
\mathbf{P}_{\mathcal{S}}=\mathbf{I}_{M}-\V\left(\V^{\rm H}\V\right)^{-1}\V^{\rm H}.
\]
The projection onto $\mathcal{T}$ is achieved by the element-wise phase normalization operation: $\boldsymbol{\Pi}_{\mathcal{T}}(\x)=e^{\jmath\angle(\x)}$. A simple scheme solving $\mathcal{P}_3$ alternates between the subspace projection onto $\mathcal{S}$ and the unit-modulus projection onto $\mathcal{T}$~\cite{bauschke1996projection,censor2012effectiveness}. This procedure is summarized in Algorithm~1, where $I_{\text{max}}$ is the maximum number of iterations, $\varepsilon_{\text{null}}$ is the nulling tolerance, and $\gamma\in(0,1)$ is a relaxation parameter that, at each step, nudges the iterate toward the decoy vector.

The per-iteration cost of Algorithm~1 under the assumption $K\!\ll\!M$ is $\mathcal{O}(MK)$, including the cost $\mathcal{O}(K^3)$ to compute the inverse of $\V^{\rm H}\V$ that happens only once.
\begin{algorithm}[t] %% be careful with the symbols' fonts
\caption{Alternating Projection for Solving $\mathcal{P}_3$}
\label{alg:pocs}
\begin{algorithmic}[1]
\STATE \textbf{Input:} $\w, \V, \gamma$, $I_{\text{max}}$, and $\varepsilon_{\text{null}}$.
\STATE Compute $\mathbf{P}_{\mathcal{S}}=\mathbf{I}_{M}-\V\left(\V^{\rm H}\V\right)^{-1}\V^{\rm H}$.
\STATE Initialize $\x^{(0)}=e^{j\angle\left(\w\right)}$.
\FOR{$t=0,1,\ldots,I_{\text{max}}-1$}
    \STATE $\bm u \leftarrow (1-\gamma)\mathbf{x}^{(t)} + \gamma \w$. \hfill
    \STATE $\bm u \leftarrow \boldsymbol{\Pi}_{\mathcal{T}}(\bm u)$. \hfill 
    \STATE $\bm u \leftarrow \mathbf{P}_{\mathcal{S}}\bm u$. \hfill 
    \STATE $\mathbf{x}^{(t+1)} \leftarrow \Pi_{\mathcal{T}}(\bm u)$. \hfill 
    \IF{$\|\V^{\rm H}\mathbf{x}^{(t+1)}\|^2\leq\varepsilon_{\text{null}}$}
        \STATE Sufficient nulling achieved; stop.
    \ENDIF
\ENDFOR
\STATE \textbf{Return:} $\boldsymbol{\omega}^{\star}=\mathbf{x}^{(I_{\text{max}})}$.
\end{algorithmic}
\end{algorithm}

\section{Analysis of the RIS-Based Decoy Capability}
\label{app:feasibility}
According to the assumptions in Section~\ref{Sec: Problem}, the feasible solution set $\mathcal{F}$ is compact and $\mathcal{P}_3$'s objective is continuous. Hence, based on the extreme value theorem~\cite{russell2020principles} and~\cite{bauschke1996projection}, a maximizer $\boldsymbol{\omega}\in\mathcal{F}$ is feasible. In addition, when $M\to\infty$ with fixed $\Delta$, $J(\theta_{\text{true}};\boldsymbol{\omega})\to 0$ holds, yielding $\text{CRB}(\theta_{\text{true}};\boldsymbol{\omega})\to\infty$. In other words, any locally unbiased AoA estimator at the radar BS becomes uninformative for the true UE direction when the RIS enforces a sufficiently deep null at it. In practice, however, $|\bar{\beta}(\theta_{\text{true}};\boldsymbol{\omega})|$ becomes small but nonzero, so the CRB remains finite yet very large.

In this section, we study the conditions for $\theta_{\text{fake}}$ to effectively conceal UE's position. Even if the hard-null constraint is satisfied when $M\!\to\!\infty$, in the finite $M$-value regime, a criterion to quantify the level of deception synthesized by the RIS is missing. To this end, we henceforth: \textit{i}) introduce a novel point-wise deception criterion % be sure that everywhere we mention to it as a criterion
and \textit{ii}) develop the corresponding set-wise version over $\boldsymbol{\Theta}_{\text{null}}$, as well as \textit{iii}) introduce a decoy-placement score for principled selection of $\theta_{\text{fake}}$.

\subsection{The $\rho$-Deception Criterion}
Using the closed-form CRB expression in~\eqref{eq: closed_form}, we define the \emph{$\rho$-deception} criterion with $\rho>1$ as a function of the RIS phase configuration vector $\boldsymbol{\omega}$  quantifying the achievable deception level with the decoy AoA $\theta_{\text{fake}}$ when the true angle is $\theta_{\text{true}}$:
\begin{equation}\label{eq: criterion}
  \frac{\mathrm{CRB}(\theta_{\text{true}};\boldsymbol{\omega})}{\mathrm{CRB}(\theta_{\text{fake}};\boldsymbol{\omega})}\ \ge\ \rho.
\end{equation}
After some manipulations and using the definition $\kappa(\theta)\triangleq
\pi^2 \cos^2(\theta)N^2(N-1)^2$, the latter criterion simplifies to:
\begin{equation}
  \left|\frac{\bar{\beta}(\theta_{\text{true}};\boldsymbol{\omega})}{\bar{\beta}(\theta_{\text{fake}};\boldsymbol{\omega})}\right|
  \ \le\ \sqrt{\frac{\kappa(\theta_{\text{fake}})}{\rho\,\kappa(\theta_{\text{true}})}}.
  \label{eq:rho-pointwise}
\end{equation}
Clearly, if $\kappa(\theta_{\text{fake}})\approx \kappa(\theta_{\text{true}})$, as is typical when $\theta_{\text{fake}}$ lies close to $\theta_{\text{true}}$, \eqref{eq:rho-pointwise}'s right-hand term reduces to $1/\sqrt{\rho}$.

%\subsection{ Band-Robust $\rho$-Deception Criterion}\label{eq: rho_band_metric}
For any discrete set of nulling angles $\boldsymbol{\Theta}_{\text{null}}$, we define:
\begin{equation}
  L_{\text{true}}(\boldsymbol{\omega})
  \triangleq \sup_{\theta\in\boldsymbol{\Theta}_{\text{null}}} \bigl|\bar{\beta}(\theta;\boldsymbol{\omega})\bigr|,
  \qquad
  \kappa_{\min} \triangleq \inf_{\theta\in\boldsymbol{\Theta}_{\text{null}}}\kappa(\theta).
\end{equation}
denoting respectively the worst-case leakage and the worst (i.e., smallest) array/illumination factor within $\boldsymbol{\Theta}_{\text{null}}$. It can be concluded that, if the following condition holds:
\begin{equation}
  \frac{L_{\rm true}(\boldsymbol{\omega})}{|\bar{\beta}(\theta_{\text{fake}};\boldsymbol{\omega})|}\ \le\ \sqrt{\frac{\kappa(\theta_{\text{fake}})}{\rho\,\kappa_{\rm min}}},
  \label{eq:rho-band1}
\end{equation}
then $\text{CRB}(\theta;\boldsymbol{\omega})\ge \rho\text{CRB}(\theta_{\text{fake}};\boldsymbol{\omega})$ $\forall\theta\in\boldsymbol{\Theta}_{\text{null}}$. This condition follows from~\eqref{eq: criterion} using $|\bar{\beta}(\theta;\boldsymbol{\omega})|\le L_{\text{true}}(\boldsymbol{\omega})$
and $\kappa(\theta)\ge \kappa_{\min}$, $\forall\theta\in\boldsymbol{\Theta}_{\text{null}}$.
Intuitively, imposing $\bar{\beta}(\theta_k;\boldsymbol{\omega})\approx 0$ $\forall k=1,\ldots,K$
reduces the worst-case leakage $L_{\text{true}}(\boldsymbol{\omega})$ up to the RIS angular resolution.
All in all, the latter inequality showcases the trade-off among the required leakage depth, the decoy angle amplitude, and the deception factor $\rho$. Analogous to the point-wise case, \eqref{eq:rho-band1} reduces to:
\begin{equation}
\sup_{\theta\in\boldsymbol{\Theta}_{\text{null}}}
\frac{|\bar{\beta}(\theta;\boldsymbol{\omega})|}{|\bar{\beta}(\theta_{\text{fake}};\boldsymbol{\omega})|}
\;\lesssim\; \frac{1}{\sqrt{\rho}}
\end{equation}
when the array/illumination factor is essentially flat over the nulling range, i.e.,
$\kappa_{\min}\approx\kappa(\theta_{\text{true}})$ (equivalently,
$\kappa(\theta)\approx\kappa(\theta_{\text{true}})$ $\forall\theta\in\boldsymbol{\Theta}_{\text{null}}$).

\subsection{Decoy Angle Selection Score}
Since $\mathcal{P}_3$'s the optimizer $\boldsymbol{\omega}$ lies in the null subspace $\mathcal{S}$ and
$\mathbf{P}_{\mathcal{S}}$ is the orthogonal projector onto $\mathcal{S}$, it holds:
\begin{align}
\mathbf{w}^{\rm H}\boldsymbol{\omega}
= \left(\mathbf{P}_{\mathcal{S}}\mathbf{w}\right)^{\rm H}\boldsymbol{\omega}
  + \underbrace{\left(\left(\mathbf{I}-\mathbf{P}_{\mathcal{S}}\right)\mathbf{w}\right)^{\rm H}\boldsymbol{\omega}}_{=0},
\end{align}
hence, using the Cauchy-Schwarz inequality, yields:
\begin{align}\label{eq: decoy_gain}
    \nonumber|\mathbf{w}^{\rm H}\boldsymbol{\omega}|&=|\left(\P_{\mathcal{S}}\mathbf{w}\right)^{\rm H}\boldsymbol{\omega}|\leq \left\|\P_{\mathcal{S}}\mathbf{w}\right\|\|\boldsymbol{\omega}\|=\left\|\P_{\mathcal{S}}\mathbf{w}\right\|\sqrt{M}\\&=M\eta(\theta_{\text{fake}}),
\end{align}
where $\eta(\theta)\triangleq \|\mathbf{P}_{\mathcal{S}}\mathbf{w}\|/\sqrt{M}$. Combining the latter expression with the angle-range-robust criterion in~\eqref{eq:rho-band1}, results in:
\begin{equation}\label{eq:UB}
\rho \;\le\; \frac{M^{2}\,\eta^{2}(\theta_{\text{fake}})\,\kappa(\theta_{\text{fake}})}
{\kappa_{\min}\,L_{\text{true}}(\boldsymbol{\omega})^{2}}
\;=\;
\frac{M^{2}\,\Phi^{2}(\theta_{\text{fake}})}{\kappa_{\min}\,L^{2}_{\text{true}}(\boldsymbol{\omega})},
\end{equation}
where $\Phi(\theta)\triangleq \eta(\theta)\,\sqrt{\kappa(\theta)}$. It is noted that a similar analysis can be done with respect to the point-wise criterion in \eqref{eq:rho-pointwise}, which is omitted here for brievity.

Expression~\eqref{eq:UB} depends on $\boldsymbol{\omega}$ through $L_{\text{true}}(\boldsymbol{\omega})$.
Moreover, under the feasibility conditions of $\mathcal{P}_3$ (i.e., $\text{rank}(\V)=K$,
$M\ge 2K$, and $\mathbf{w}\notin\operatorname{span}(\mathbf{V})$), hard nulls are achievable and
$L_{\text{true}}(\boldsymbol{\omega})$ can be made arbitrarily small, hence, there is no finite upper bound on $\rho$ that holds uniformly over all $\boldsymbol{\omega}\in\mathcal{F}$. As the nulls deepen, the CRB at the true angle $\theta_{\text{true}}$ diverges. A configuration-independent upper bound is possible by imposing a design leakage cap, i.e., if $L_{\text{true}}(\boldsymbol{\omega})\ \le\ \overline{L}$ holds for all feasible $\boldsymbol{\omega}$, then:
\begin{equation}\label{eq: rho_UB}
\rho\ \le\ \rho_{\text{UB}}(\theta_{\text{fake}};\Bar{L})\triangleq\frac{M^{2}\,\Phi^{2}(\theta_{\text{fake}})}{\kappa_{\min}\,\overline{L}^{\,2}}.
\end{equation}
The maximization of this upper bound over admissible decoy angles, which can be expressed as follows:
\[
\theta_{\text{fake}}^{\star}
\;=\;
\underset{\theta\notin\boldsymbol{\Theta}_{\text{null}}}{\arg\max}\;
\rho_{\text{UB}}\left(\theta;\overline{L}\right),
\]
provides a principled decoy AoA selection criterion. Intuitively, we prefer angles at which the RIS can radiate strongly (i.e., large $\eta(\theta)$) and the BS becomes highly directive at (i.e., large $\kappa(\theta)$). Note, however, that maximizing $\Phi(\theta)$ and thus only the upper bound guarantees a large CRB contrast between the decoy and the nulled region; however, the optimal $\rho$-deception criterion is not ensured. It is mentioned that the realized $\rho$ also depends on the decoy gain actually achieved at that angle as well as the
residual leakage attained by the optimization, both of which may vary with $\theta_{\text{fake}}$.

In practice, one could shortlist the top candidate decoy AoAs with the largest $\rho_{\text{UB}}\left(\theta;\overline{L}\right)$, solve the optimization for each candidate angle, compute the realized deception factor $\rho$, and select the AoA that delivers the best deception level.
\begin{figure*}[!t]
  \begin{subfigure}[t]{0.45\textwidth}
  \centering
    \includegraphics[width=\textwidth]{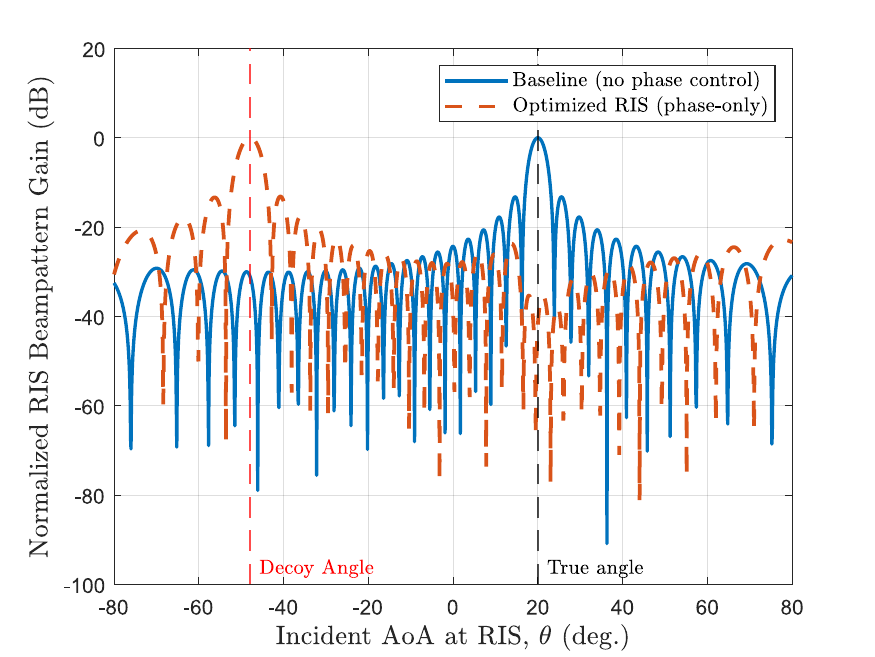}
    \caption{Normalized RIS beampattern gain.}
    \label{fig:beam}
  \end{subfigure}\hfill
  \begin{subfigure}[t]{0.45\textwidth}
  \centering
    \includegraphics[width=\textwidth]{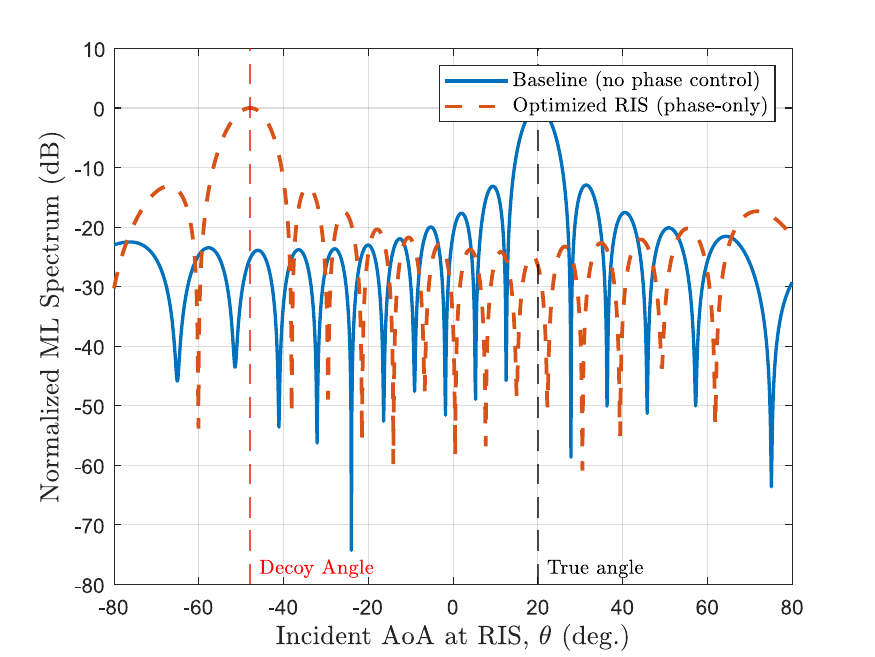}
    \caption{ML AoA spectrum.}
    \label{fig:ML}
  \end{subfigure}
  \caption{\small{Normalized RIS beampattern gain (left) and corresponding ML-based AoA spectrum (right) considering a radar BS with \(N=16\) antennas and an RIS with \(M=32\) elements. \(K=10\) nulling angles have been imposed over a \(\Delta=3^{\circ}\) angle window; the decoy and true AoAs were \(\theta_{\text{fake}}=-48^{\circ}\) and \(\theta_{\text{true}}=20^{\circ}\), respectively.}}\vspace{-0.4cm}
  \label{fig:beam_ML} %what do you mean with this phase-only?
\end{figure*}

\begin{figure}[!t]
	\begin{center}
	\includegraphics[scale=0.6]{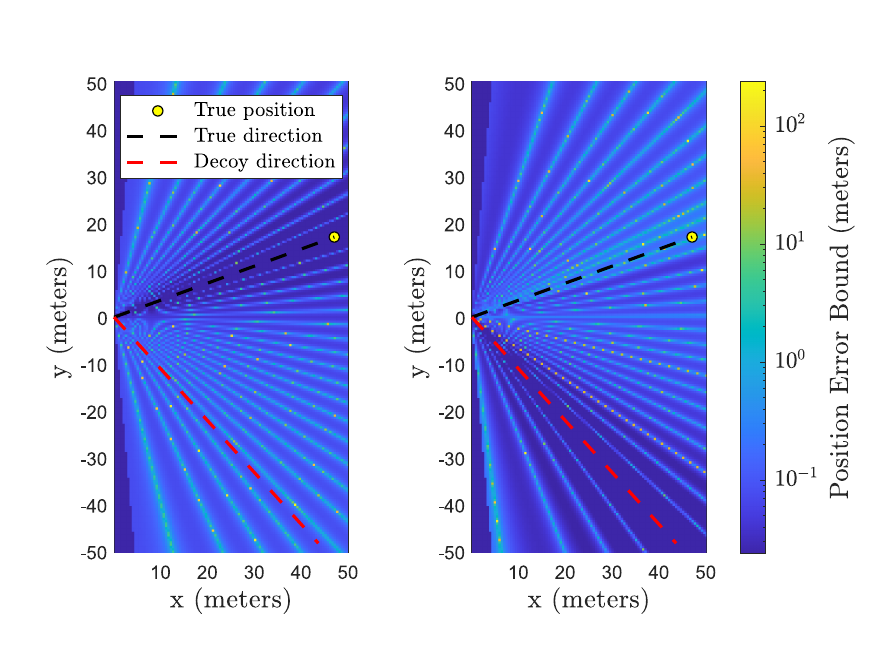}
	\caption{\small{PEB heatmaps before (left) and after (right) the proposed RIS-enabled spoofing using the parameters of Fig.~\ref{fig:beam_ML}. Before optimization, the monostatic radar exhibits a sharp PEB with a minimum at the true AoA, whereas, after optimization, the PEB minimum shifts to the prescribed decoy angle, while the true angle is obfuscated without any distinct minimum remaining at the actual UE direction.}} 
    \vspace{-0.4cm}
	\label{fig: PEBfig}
	\end{center}
\end{figure}

\section{Numerical Results and Discussion}\label{Sec: results}
In this section, we present numerical results for the previously presented decoy-angle synthesis design and assess the theoretical and actual deception performance via the \emph{$\rho$-deception} criterion. A simulation setup with the following parameters was implemented: $20$ GHz carrier frequency with a coherent block span of $T=50$ pilot transmissions, a radar BS with $N=16$ antennas, and an RIS-coated UE with $M=32$ reflecting elements positioned at $\p_{\rm RIS}=[48,17]$ meters. Unless stated otherwise, the decoy AoA was \(\theta_{\text{fake}}=-48^{\circ}\), and given \(\p_{\rm RIS}\), the true AoA was \(\theta_{\text{true}}=20^{\circ}\). The AWGN power was set to \(\sigma^2=-80\) dBm, the nulling window width to \(\Delta=3^{\circ}\) with \(K=10\) uniformly spaced null directions, and the transmit power to \(P=20\) dBm.

After collecting $T$ pilot symbols, the BS performed ML estimation of the AoA $\theta_{\text{true}}$. Let the sample spatial covariance matrix be $\mathbf{R} = \frac{1}{T}\Y\Y^{\rm H}$. Concentrating the likelihood over the unknown complex gain yields the standard objective:
\begin{equation}
\theta_{\text{true}}^{\star}
= \arg\max_{\theta} \bigl|\boldsymbol{\alpha}_N^{\rm H}(\theta)\mathbf{R}\boldsymbol{\alpha}_N(\theta)\bigr|^{2}.
\end{equation}
Structurally, the ML form is unchanged by the RIS, which can only shape the angular gain
$|\bar{\beta}(\theta_{\text{true}};\boldsymbol{\omega})|$. As shown in Section~\ref{Sec: 3}, the AoA FI scales with $|\bar{\beta}(\theta_{\text{true}};\boldsymbol{\omega})|^{2}$. Thus, if the RIS imposes a deep null at the true direction, the FI collapses there and the AoA CRB diverges. This implies that any unbiased AoA estimator at the radar BS, including ML, becomes extremely unreliable at~\(\theta_{\text{true}}\).

\subsection{Position Estimation Performance}
Figure~\ref{fig:beam_ML} contrasts the RIS beampattern gain and the corresponding ML-based AoA spectrum before and after the proposed RIS-aided spoofing. As seen, with no phase control (i.e., \([\boldsymbol{\omega}]_m=1 \forall m\)), the beampattern returns energy toward the true UE direction \(\theta_{\text{true}}\). However, after the proposed RIS optimization, energy is redirected to the prescribed decoy AoA while a deep, banded null is carved around~\(\theta_{\text{true}}\). The ML spectrum mirrors this redistribution: it is flat and uninformative near~\(\theta_{\text{true}}\) exhibiting a sharp peak at the decoy AoA, systematically pulling the estimator away from the actual target. This behavior aligns with the FI/CRB analysis: suppressing kernel power at~\(\theta_{\text{true}}\) collapses local FI, inflating the CRB, whereas concentrating power at the decoy tightens the bound and stabilizes the ML peak. The banded null enhances robustness to small angular mismatches, and subdued sidelobes limit leakage that might otherwise reveal~\(\theta_{\text{true}}\).

Figure~\ref{fig: PEBfig} visualizes the PEB before and after applying our RIS-enabled spoofing. The deception effect is evident. Prior to optimization, the heatmap exhibits a sharp, localized minimum at~\(\theta_{\text{true}}\), indicating high estimation precision there. On the contrary, after optimization, that valley disappears and the global minimum relocates to the decoy AoA site, while the vicinity of~\(\theta_{\text{true}}\) becomes uniformly high-PEB dominated. This migration of the PEB minimum from~\(\theta_{\text{true}}\) to~\(\theta_{\text{fake}}\) is the spatial counterpart of the angular manipulation in Fig.~\ref{fig:beam_ML}. By imposing a deep banded null at~\(\theta_{\text{true}}\) and concentrating kernel power at the decoy AoA, the RIS increases the CRB at the former direction and reduces it at the latter, thereby steering locally unbiased estimators away from the UE.

\subsection{Investigation of the \emph{$\rho$}-Deception Criterion}
\begin{figure}[!t] %remove the text "Leakage radio threshold"' from the legend, keep only the values without parentheses
	\begin{center}
	\includegraphics[scale=0.60]{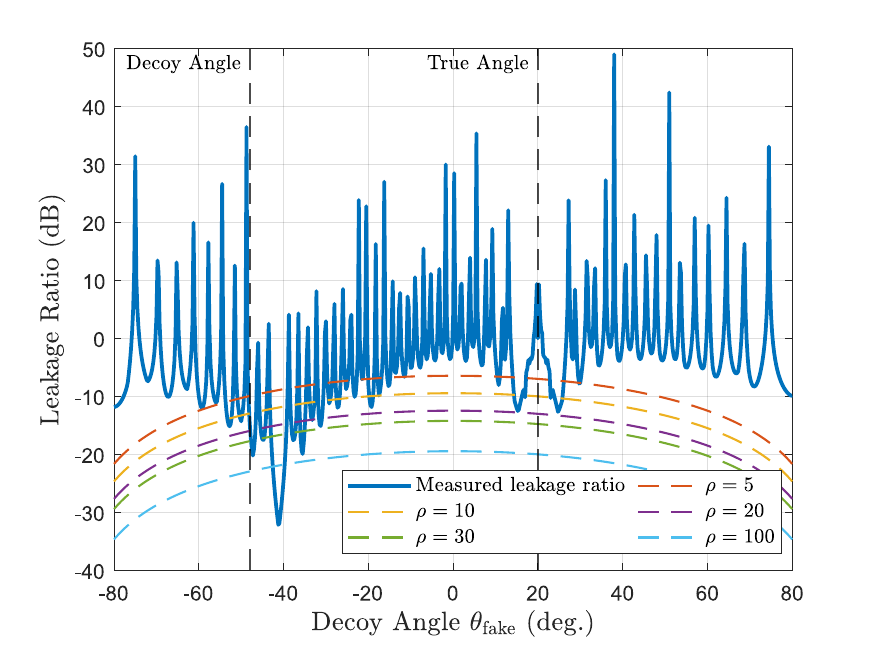}
	\caption{\small{Leakage ratio \(L_{\rm true}(\boldsymbol{\omega})/|\bar{\beta}(\theta_{\text{fake}};\boldsymbol{\omega})|\) versus the decoy angle $\theta_{\text{fake}}$ including the thresholds \(\sqrt{\kappa(\theta_{\text{fake}})/(\rho\kappa_{\min})}\) for various \(\rho\)-deception levels, using the optimized RIS phase profile of Fig.~\ref{fig:beam_ML}. It is shown that, although the design yields low leakage at \(\theta_{\text{fake}}=-48^{\circ}\), it is not globally optimal, as the hard-null constraint prioritizes enforcing a banded null around the angle \(\theta_{\text{true}}=20^{\circ}\).}} 
    \vspace{-0.4cm}
	\label{fig: ratio}
	\end{center}
\end{figure}

Figure~\ref{fig: ratio} illustrates the trade-off between~\(\theta_{\text{fake}}\)'s amplitude and residual leakage, by plotting the leakage ratio \(L_{\rm true}(\boldsymbol{\omega})/|\bar{\beta}(\theta_{\text{fake}};\boldsymbol{\omega})|\) versus the decoy angle, together with band-robust \(\rho\)-deception thresholds \(\sqrt{\kappa(\theta_{\text{fake}})/(\rho\,\kappa_{\min})}\) for the optimized RIS-based spoofing design of Fig.~\ref{fig:beam_ML}. Decoy angles where the curve falls below a threshold achieve the corresponding \(\rho\)-deception level; those above fail it. It is also shown that the optimized RIS profile attains a low ratio at the chosen decoy (here \(\theta_{\text{fake}}=-48^\circ\)), evidencing effective deception, yet it is not the global minimum. This behavior reflects the priority of enforcing a hard, banded null around \(\theta_{\text{true}}=20^\circ\), which necessarily limits decoy gain and elevates worst-case leakage. Consistent with Fig.~\ref{fig: PEBfig}, the beam toward the decoy position is relatively broad. It can, thus, be concluded from Fig.~\ref{fig: ratio} that the proposed spoofing design provides deeper, band-robust nulls enhancing concealment around~\(\theta_{\text{true}}\), but it limits the achievable decoy gain, with the realized strength governed by the efficiency of the solution of the $\mathcal{P}_3$ problem.
\begin{figure}[!t] %y-axis: Upper Bound $\rho_{\text{UB}}$
	\begin{center}
	\includegraphics[scale=0.6]{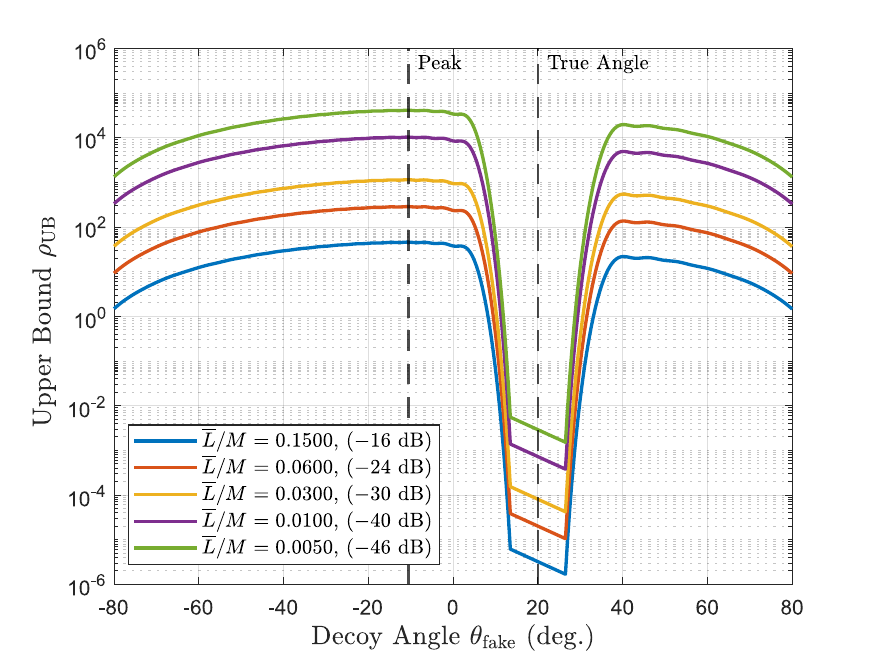}
	\caption{\small{The upper bound $\rho_{\text{UB}}$ versus the decoy AoA $\theta_{\text{fake}}$ for several leakage caps \(\overline{L}\). As observed, varying \(\overline{L}\) scales the bound vertically without shifting the peak. The plot also delineates the nulled window around \(\theta_{\text{true}}\) where deception is unattainable from the region of the feasible decoy AoAs.
}} 
    \vspace{-0.4cm}
	\label{fig: uppper}
	\end{center}
\end{figure}

Finally, the RIS-agnostic upper bound~\(\rho_{\mathrm{UB}}\left(\theta; \overline{L}\right)\) in~\eqref{eq: rho_UB} is evaluated with respect to the decoy angle~$\theta_{\text{fake}}$ in Fig.~\ref{fig: uppper}. It can be seen that the peak locations are governed by \(\Phi(\theta)\): they occur when the RIS can radiate strongly after projection onto the true AoA null subspace (i.e., large \(\eta(\theta)\)) and the radar BS array/illumination factor is highly directive (i.e., large \(\kappa(\theta)\)). As observed, varying \(\overline{L}\), merely rescales the curve without shifting these maxima: the same angles remain optimal targets for deception irrespective of the leakage budget, while the pronounced dip across the nulled band around \(\theta_{\text{true}}\) confirms that robust nulling renders deception unattainable there. % check this sentence

\section{Conclusion}
In this paper, we studied RIS-enabled spoofing against adversary monostatic sensing. We introduced a compact RIS kernel that links the phase profile to the adversary’s angular response and showed, via FI and CRB analyses, that AoA accuracy scales with kernel power. %Thus, deepening a kernel null at the true bearing inflates the CRB locally, while concentrating power at a decoy tightens it.
Leveraging this insight, we formulated an RIS design problem formulation that maximizes decoy gain under hard, band-robust nulls around the true AoA. We further quantified deception via point- and set-wise \(\rho\)-criteria, and proposed a configuration-agnostic AoA decoy selection score derived from an analytic upper bound. The presented numerical results confirmed deep true AoA nulls and a pronounced decoy peak, with the latter rendering ML AoA estimates unreliable for the RIS-coated target delineating the practical limits of the proposed RIS-enabled spoofing design.

\bibliographystyle{IEEEtran}

\bibliography{ms}

\begin{thebibliography}{10}
\providecommand{\url}[1]{#1}
\csname url@samestyle\endcsname
\providecommand{\newblock}{\relax}
\providecommand{\bibinfo}[2]{#2}
\providecommand{\BIBentrySTDinterwordspacing}{\spaceskip=0pt\relax}
\providecommand{\BIBentryALTinterwordstretchfactor}{4}
\providecommand{\BIBentryALTinterwordspacing}{\spaceskip=\fontdimen2\font plus
\BIBentryALTinterwordstretchfactor\fontdimen3\font minus \fontdimen4\font\relax}
\providecommand{\BIBforeignlanguage}[2]{{%
\expandafter\ifx\csname l@#1\endcsname\relax
\typeout{** WARNING: IEEEtran.bst: No hyphenation pattern has been}%
\typeout{** loaded for the language `#1'. Using the pattern for}%
\typeout{** the default language instead.}%
\else
\language=\csname l@#1\endcsname
\fi
#2}}
\providecommand{\BIBdecl}{\relax}
\BIBdecl

\bibitem{huang2019reconfigurable}
C.~Huang \emph{et~al.}, ``Reconfigurable intelligent surfaces for energy efficiency in wireless communication,'' \emph{IEEE Trans. Wireless Commun.}, vol.~18, no.~8, pp. 4157--4170, 2019.

\bibitem{9367575}
G.~C. Alexandropoulos \emph{et~al.}, ``Phase configuration learning in wireless networks with multiple reconfigurable intelligent surfaces,'' in \emph{Proc. IEEE Global Commun. Conf.}, Taipei, Taiwan, 2020.

\bibitem{alexandropoulos2023ris}
------, ``{RIS}-enabled smart wireless environments: {D}eployment scenarios, network architecture, bandwidth and area of influence,'' \emph{EURASIP J. Wireless Commun. and Netw.}, vol. 2023, no.~1, p. 103, 2023.

\bibitem{alexandropoulos2021reconfigurable}
------, ``Reconfigurable intelligent surfaces for rich scattering wireless communications: {R}ecent experiments, challenges, and opportunities,'' \emph{IEEE Commun. Mag.}, vol.~59, no.~6, pp. 28--34, 2021.

\bibitem{RIS_massive_access}
X.~Cao \emph{et~al.}, ``Massive access of static and mobile users via reconfigurable intelligent surfaces: Protocol design and performance analysis,'' \emph{IEEE J. Sel. Areas Commun.}, vol.~40, no.~4, pp. 1253--1269, 2022.

\bibitem{RIS_smart_cities}
H.~Chen \emph{et~al.}, ``{RISs} and sidelink communications in smart cities: The key to seamless localization and sensing,'' \emph{IEEE Commun. Mag.}, vol.~61, no.~8, pp. 140--146, 2023.

\bibitem{liu2025sustainable}
R.~Liu \emph{et~al.}, ``Sustainable wireless networks via reconfigurable intelligent surfaces ({RIS}s): Overview of the {ETSI ISG RIS},'' \emph{IEEE Commun. Standards Mag.}, 2025.

\bibitem{3gppRel20}
{3rd Generation Partnership Project (3GPP)}, ``Release 20,'' 3GPP Specifications \& Technologies, 2027.

\bibitem{Secure_ISAC}
N.~Su \emph{et~al.}, ``Integrating sensing and communications in {6G}? {N}ot until it is secure to do so,'' \emph{arXiv preprint arXiv:2503.15243}, 2025.

\bibitem{zheng2024intelligent}
B.~Zheng \emph{et~al.}, ``Intelligent reflecting surface-aided electromagnetic stealth against radar detection,'' \emph{IEEE Trans. Signal Process.}, vol.~72, pp. 3438--3452, 2024.

\bibitem{wang2024intelligent}
H.~Wang \emph{et~al.}, ``Intelligent reflecting surface-aided radar spoofing,'' \emph{IEEE Wireless Commun. Lett.}, 2024.

\bibitem{magbool2025hiding}
A.~Magbool \emph{et~al.}, ``Hiding in plain sight: {RIS}-aided target obfuscation in {ISAC},'' \emph{arXiv preprint arXiv:2503.05418}, 2025.

\bibitem{RIS_blinding}
G.~Rexhepi \emph{et~al.}, ``Blinding the wiretapper: {RIS}-aided user occultation in the {ISAC} era,'' in \emph{Proc. {Asilomar} Signals, Sys. Comp. Conf.}, Pacific Grove, USA, 2025.

\bibitem{6184250}
M.~Matthaiou \emph{et~al.}, ``Analytic framework for the effective rate of {MISO} fading channels,'' \emph{IEEE Trans. Commun.}, vol.~60, no.~6, pp. 1741--1751, 2012.

\bibitem{kay1993statistical}
S.~M. Kay, ``Statistical signal processing: estimation theory,'' \emph{Prentice Hall}, 1993.

\bibitem{van2002optimum}
H.~L. Van~Trees, ``Optimum array processing: {P}art {IV} of detection, estimation, and modulation theory,'' \emph{John Wiley \& Sons}, 2002.

\bibitem{shen2010fundamental}
Y.~Shen and M.~Z. Win, ``Fundamental limits of wideband localization—{P}art {I}: {A} general framework,'' \emph{IEEE Trans. Inf. Theory}, vol.~56, no.~10, pp. 4956--4980, 2010.

\bibitem{cox1987robust}
H.~Cox \emph{et~al.}, ``Robust adaptive beamforming,'' \emph{IEEE Trans. Acoust., Speech, Signal Process.}, vol.~35, no.~10, pp. 1365--1376, 1987.

\bibitem{russell2020principles}
B.~Russell, ``Principles of mathematics,'' \emph{Routledge}, 2020.

\bibitem{bauschke1996projection}
H.~H. Bauschke and J.~M. Borwein, ``On projection algorithms for solving convex feasibility problems,'' \emph{SIAM}, vol.~38, no.~3, pp. 367--426, 1996.

\bibitem{censor2012effectiveness}
Y.~Censor \emph{et~al.}, ``On the effectiveness of projection methods for convex feasibility problems with linear inequality constraints,'' \emph{Comput. Optimization Appl.}, vol.~51, no.~3, pp. 1065--1088, 2012.

\bibitem{dinkelbach1967nonlinear}
W.~Dinkelbach, ``On nonlinear fractional programming,'' \emph{Manag. Sci.}, vol.~13, no.~7, pp. 492--498, 1967.

\bibitem{boyd2004convex}
S.~P. Boyd and L.~Vandenberghe, ``Convex optimization,'' \emph{Cambridge University Press}, 2004.

\end{thebibliography}

\end{document}